\documentclass[10pt,aps,prl,reprint,showpacs,superscriptaddress,floats,floatfix,amsmath,amssymb]{revtex4-1}
\usepackage[pdfpagemode=None]{hyperref}
\usepackage{graphicx,bm}
\usepackage{braket}
\usepackage{mathrsfs}
\usepackage{wasysym}
\usepackage{epstopdf}
\usepackage{color}
\begin{document}

\title{Hong-Ou-Mandel Interference with a Single Atom}

\author{K.A. Ralley}
\affiliation {University of Birmingham, School of Physics \& Astronomy, B15 2TT, UK}

\author{I.V. Lerner }
\affiliation {University of Birmingham, School of Physics \& Astronomy, B15 2TT, UK}
\author{I.V. Yurkevich *}
 \affiliation{ Aston University, School of Engineering \& Applied Science, Birmingham, B4 7ET, UK
}

\begin{abstract}{The Hong-Ou-Mandel (HOM) effect is widely regarded as the quintessential quantum interference phenomenon in optics. In this work we examine how nonlinearity can smear statistical photon bunching in the HOM interferometer. We model both the nonlinearity and a balanced beam splitter with  a single two-level system and calculate a finite probability of anti-bunching arising in this geometry. We thus argue that the presence of such nonlinearity would reduce the visibility in the standard HOM setup, offering  some explanation for the diminution of the HOM visibility observed in many experiments. We use the same model to show that the nonlinearity affects a resonant two-photon propagation through a two-level impurity in a waveguide due to a ``weak photon blockade"  caused by the impossibility of double-occupancy and argue that this effect might be stronger for multi-photon propagation.}
\end{abstract}


\maketitle

\section{Introduction}
\begin{figure}
\includegraphics[width=0.9\columnwidth]{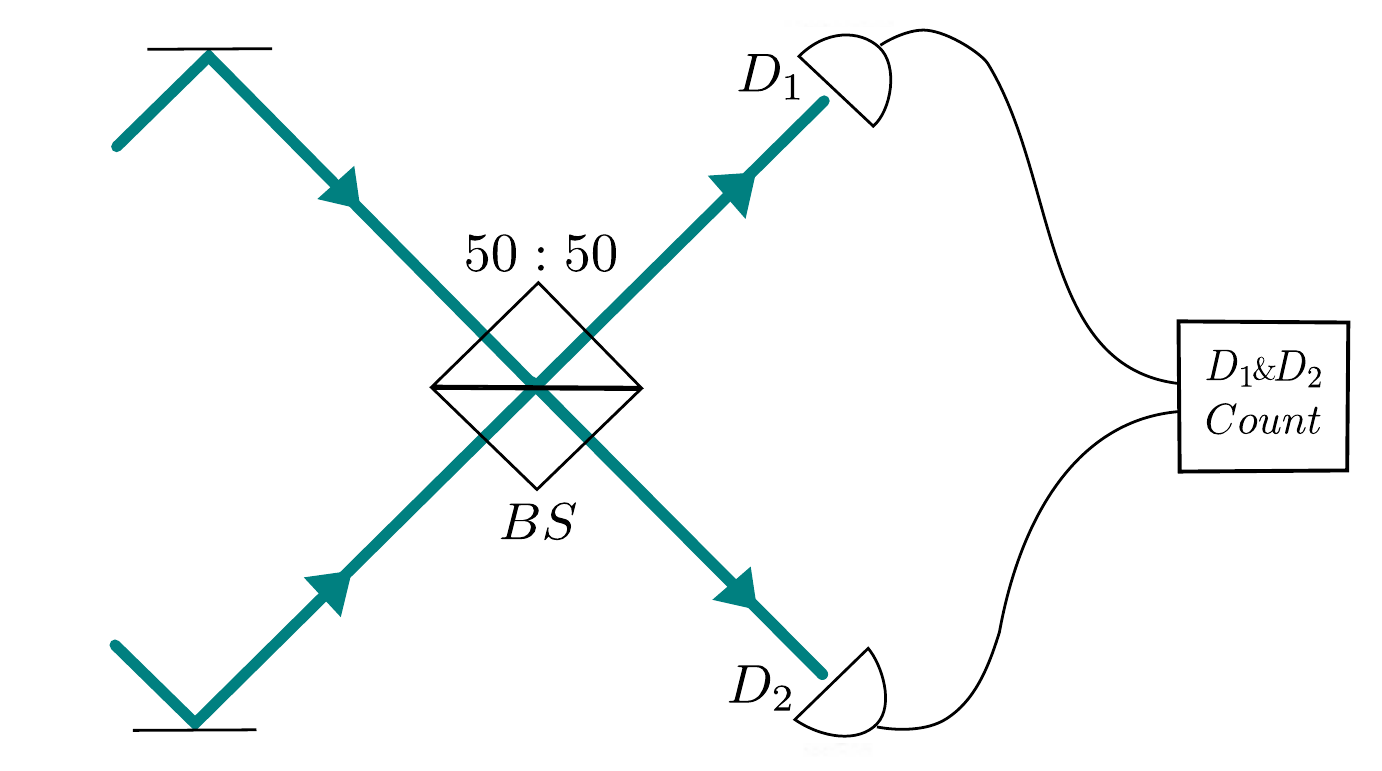}
\caption{Traditional Hong-Ou-Mandel interference scheme. Two identical photons arriving \emph{simultaneously} at a balanced, broadband beam splitter (BS) will be conveyed along only one of the possible outgoing channels - and so, in contrast to the general case, no coincidence counts will be accumulated. \label{Homi}
}
\end{figure}

The Hong-Ou-Mandel (HOM) interferometer\cite{HOM1987,Mandel1995} is one of the main tools in registering biphotons (i.e.\  entangled photon pairs) created by  SPDC (spontaneous parametric down-conversion) \cite{Ou1988,Pan2001,Kuzucu:05,Mosley2008,Eckstein:08,Nasr2008,Ray:11,Douce:13,Jin2014}.
 The ideal HOM protocol comprises a four-port  interferometer with two incoming and two outgoing channels,  a spectrally-flat, balanced beam splitter and  the coincidence counter (see Fig.~\ref{Homi}). When uncorrelated photons arrive at the incoming port, the outgoing photons split equally  between the two detectors resulting in a signal in the coincidence counter. However, when two photons arrive \emph{simultaneously} at both the incoming ports,  the outgoing photons are bunched together leaving the interferometer through only one of the outgoing ports \cite{HOM1987,Mandel1995}, so that the counter detects no signal.
As the  entangled  photons are simultaneously created by SPDC, their arrival at the two incoming ports  (against the background of uncorrelated arrivals of non-entangled photons) is manifested by a dip in a signal registered by the coincidence counter.

In this work we argue that characteristics of such a dip can be substantially affected by the presence of a nonlinearity  that leads to an effective interaction between two photons that simultaneously enter the beam splitter in the HOM device. The nonlinearity might be due to a  single two-level impurity embedded in the device. The interplay between such a nonlinearity and HOM interference leads to a novel mechanism of the suppression of the photon bunching which might obscure information about biphoton generation.

 An alternative geometry where nonlinearity is also essential is illustrated in Fig.~\ref{pc}a: a two-level impurity is embedded in a waveguide. Such an impurity would practically not affect the propagation of off-resonance photons but result in a full reflection of a single photon at the resonant frequency, acting similarly to the side-attached resonant impurity for electrons propagating through a conducting wire \cite{LYY:08}. In this geometry the nonlinearity suppresses the resonant reflection and leads to a partial propagation for the photon pair at the resonance.   A feasible ``dual"   geometry, Fig.~\ref{pc}b,  comprises two waveguides separated by an optically opaque region with an embedded two-level impurity. There is no propagation of non-resonant photons through such a weak link but an ideal resonant transmission of a single photon, which is again suppressed by the nonlinearity. We show that all such geometries can be described within essentially the same model where the nonlinearity substantially alters two-photon propagation.

\section{The model}
\begin{figure}
\includegraphics[width=0.9\columnwidth]{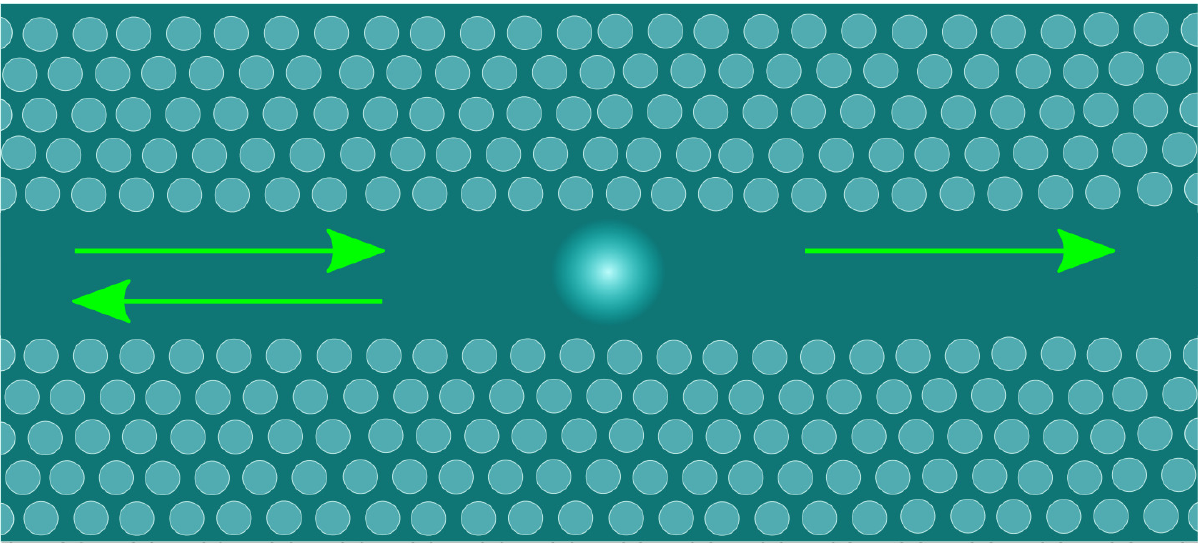}
\centerline{(a)}
 \includegraphics[width=0.9\columnwidth]{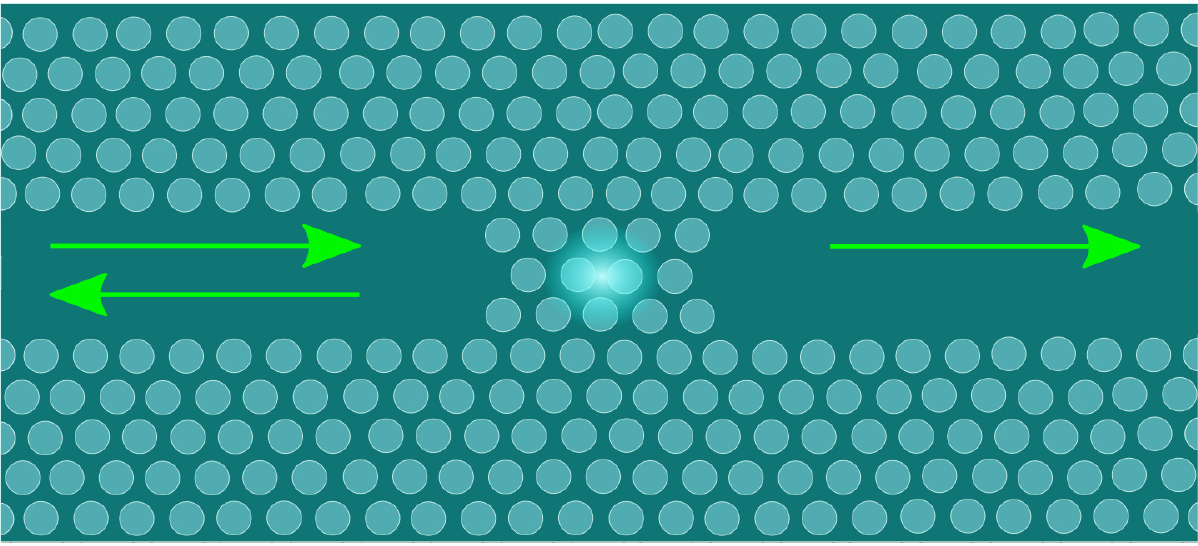}
 \centerline{(b) }
\caption{Two additional geometries described by the model of Eq.~(\ref{H}): (a) a single atom (TLS) embedded in a 1D photonic crystal waveguide leads to reflection of resonant photons in the channel (off-resonant photons are freely transmitted); (b) an interstitial TLS provides a resonant link between two waveguide channels. \label{pc}
}
\end{figure}

 We consider  a wave-guided few-photon beam interacting with a single near-resonant atom that can be described as a two-level system (TLS). We assume that both incident and transmitted or reflected photons can propagate along two channels. Such a model mimics (under conditions specified below) the HOM geometry of Fig.~\ref{Homi} when  the  incoming channels are different, and corresponds to a TLS  embedded into a 1D photonic crystal waveguide when photons enter through the same channel, Fig.~\ref{pc}.  The corresponding  Hamiltonian is reduced in the usual rotating wave approximation to
\begin{align}
\hat{H} &= \sum_{\alpha}\varint \frac{\mathrm{d} k }{2\pi } \Big[k\,\hat{b}_{\alpha,k}^{\dag} \hat{b}^{\phantom{\dagger  }}_{\alpha,k}   +\sqrt{\gamma} \big(\hat{\sigma}^{\phantom{\dagger  }}_{+}
\hat{b}^{\phantom{\dagger  }}_{\alpha, k} +\mathrm{h.c.} \big)\Big].\label{H}
\end{align}
Here $\hat{b}^+_{\alpha,k}  $ and $b^{\phantom{\dagger  }}_{\alpha,k} $ are the photon creation and annihilation operators with the index $\alpha = 1, 2$ labelling the channels; $\omega =k$ is the  photon energy (counted from the upper energy level of the TLS) in the units where both $\hbar$  and the group velocity of light in the medium are set to 1;  the TLS is described by  Pauli's rasing and lowering matrices, $\hat{\sigma} _\pm =\hat{\sigma }_x\pm  i\hat{\sigma }_y $; and $\sqrt{\gamma}$ is the atom-photon coupling strength, with $\gamma$ being the TLS relaxation rate. \emph{Neglecting} the nonlinearity, i.e.\ considering two- or multi-photon scattering from the TLS as totally independent, this model describes -- with a proper choice of $\gamma$ and $\omega$ specified below -- both a  50:50 (balanced) beam splitter for the HOM geometry and ideal resonant reflection or transmission for the geometry of Fig.~\ref{pc}(a) and (b), respectively.

We will describe how the effects of the nonlinearity caused by   indirect photon interaction via  scattering from the TLS change the probability $P_{\alpha\alpha' |\beta\beta'} $ of two  photons entering via channels $\beta$ and $\beta'$ and exiting via  channels $\alpha$ and $\alpha'$
in the \emph{Results} below.  In both cases  of the HOM geometry ($\beta\not=\beta'$) and the resonance geometry ($\beta=\beta'$) we calculate the probability of photons leaving through different channels: $P_\mathrm{HOM}=P_{12|12}+P_{21|12}  $ and $P_\mathrm{res}=P_{12|\beta\beta}+P_{21|\beta\beta}  $. The former describes the suppression of photon bunching in the HOM experiment while the latter describes a ``weak photon blockade'' of the resonant two-photon reflection from or transmission through TLS.

Before describing the nonlinearity effects, we express the probability $P_{\alpha\alpha' |\beta\beta' }^0 $ of independent two-photon scattering from the TLS via parameters of the model.
One-photon scattering from a TLS is described by the  scattering matrix $\hat{\mathcal{S}} \equiv\{{s_{\alpha\beta} }\} $ connecting incident and outgoing photons in the two channels:
\begin{align}\label{SM}
\ket{\alpha}_\mathrm{out}  = s_{\alpha\beta}\,\ket{\beta}_\mathrm{in},\quad
\hat{\mathcal{S}}=\left(
          \begin{array}{cc}
            t  & r' \\
            r  & t'  \\
          \end{array}
        \right)\,,\quad \hat{\mathcal{S}}^{\dagger}\hat{\mathcal{S}}=\hat1\,.
\end{align}
 The unitarity of S-matrix requires the equality of the reflection and transmission amplitudes,  $|t|=|t' |$ and $|r|=|r' |$. In a system with reflection symmetry the choice $r=r'$ and $t=t'$   is assumed in what follows.

For a monochromatic photon in the resonant transmission geometry of Fig.~\ref{pc}(b),  the transmission and reflection amplitudes in Eq.~(\ref{SM}) become
\begin{align}\label{sa}
r\to r_{\omega}=\frac{\omega}{\omega+i\gamma}\,,\quad t \to t_{\omega}=\frac{-i\gamma}{\omega+i\gamma}\,,
\end{align}
where $\omega $ is a photon detuning from the resonance, and the TLS is assumed to be initially in its ground state. The dual geometry of Fig.~\ref{pc}(a) is obtained by swapping the reflection and transmission amplitudes, $r\rightleftarrows t$. In both these cases (choosing the incoming channel $\beta=\beta' =1$),  $P_{\alpha\alpha' |11}^0 $ is just a product of one-photon probabilities so that $P_{11|11}^0 =1$  in  the  geometry of Fig.~\ref{pc}(a) (both the photons are resonantly reflected), or $P_{22|11}^0=1$ in the dual  geometry of Fig.~\ref{pc}(b) (both are resonantly transmitted). Only nonlinearity will lead to distinct features in two-photon scattering (as compared to single-photon) in this geometry.

 For the HOM geometry   ($\beta\ne\beta'$),   the probability of anti-bunching  without nonlinearity is
\begin{align}\label{anti-b}
P_{\mathrm{HOM}}^0=|s_{11} s_{22}+s_{12} s_{21}|^2=|t^2+r^2|^2=(T-R)^2\!,
\end{align}
 where $T=|t|^2$ and $R=|r|^2$ are the single-photon transmission and reflection probabilities. Thus a perfect bunching, $P^0_\mathrm{HOM}=0  $, occurs when  $T=R=\frac{1}{2}$ so that
  the model of Eq.~(\ref{H}) would emulate   an ideally balanced  beam splitter when the frequency  $\omega  $ of   monochromatic  incoming photons coincides with the resonance width $\gamma $. For realistic  time-resolved photons, an almost balanced beam splitter will be emulated if both photons have   spectral {functions}  centred at $\omega=\gamma$ of width $\sigma$ much smaller than $\gamma$. The limitation $\Gamma\equiv\gamma/\sigma\gtrsim1$ is essential for the model we consider here.

This result holds for two identical  photons. If the photons become distinguishable, e.g.\ by a delay $\tau$ between their arrival times, they would   be uncorrelated for $\tau \gg\sigma^{-1}  $ and thus  have equal probabilities to go to ports 1 or 2. Therefore,
one expects a dip in $P^0_\mathrm{HOM}(\tau)$ at $\tau=0$ with a non-universal shape dependent on the spectral function. When the latter is Gaussian, one has
\begin{align}
P^0_\mathrm{HOM} (\tau)=\frac{1}{2}\,\left[1- \mathrm{e}^{-(\sigma\tau)^2}\right]\,.
\end{align}
 The depth of such a  dip is thus used, e.g.,  to characterise a rate of the SPDC biphoton production, as the shape of the entangled photons is identical and they are created simultaneously.

Here we argue   that a non-ideal dip may result from the nonlinearity in the HOM beam splitter like that described by the Hamiltonian ({\ref{H}}).

\section{Results}

\begin{figure}[t]
\includegraphics[width=.92\columnwidth]{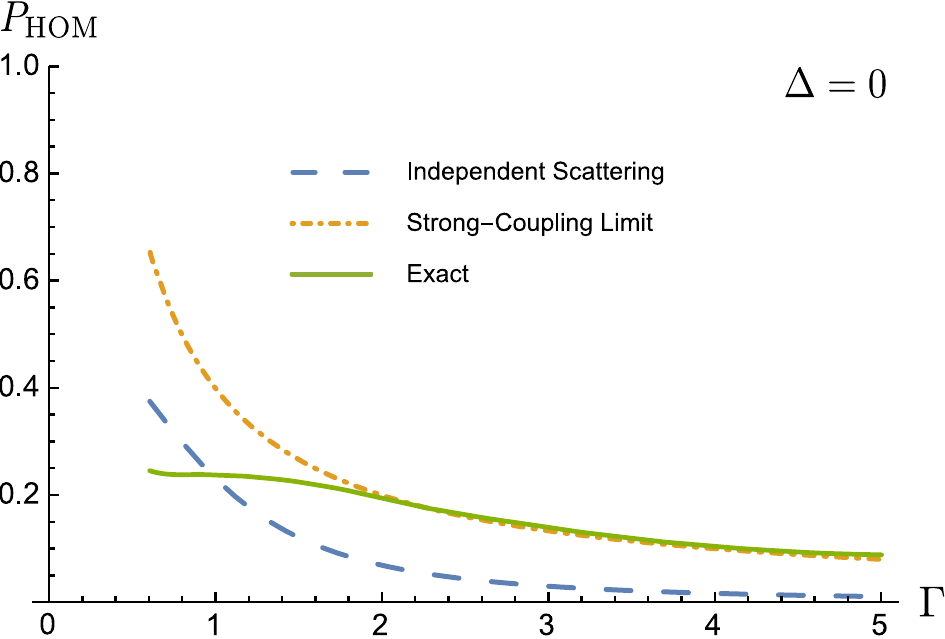}
\caption{\label{Fig4}
The anti-bunching probability in the HOM geometry. Neglecting the nonlinearity (independent scattering), it goes to $0$ with  $\Gamma\to\infty$ (the monochromatic limit), which corresponds to the balanced beam splitter.  The nonlinearity makes the probability of anti-bunching finite for any value of $\Gamma$ (the solid line describes the exact numerical solution); the analytic asymptotics (the strong-coupling -- monochromatic -- limit) practically coincides with the exact solution for $\Gamma\gtrsim2$.  }
\end{figure}
 In the HOM geometry,  when the identical photons come from the opposite ($\beta \ne\beta '$) channels without a time delay, the nonlinearity results in a nonzero probability $P _\mathrm{HOM}({\tau\ne0}) $ of detecting the two photons in different outgoing channels. We find the probability of the coincidence counter clicking as a function of the time delay $\tau$ between two photons hitting the TLS as
\begin{align}\label{P12}
P_{\mathrm{HOM}}({\tau })=P^0_{\mathrm{HOM} }({\tau })+ \delta P _\mathrm{HOM} (\tau)\,,
\end{align}
where the last term corresponds to the nonlinearity-induced photon-photon interaction.  In particular,  for photons with narrow  Gaussian spectral function \cite{note1}, when $\Gamma \gg1$,  we find
 \begin{align}\label{HOMres}
     \delta P _\mathrm{HOM} (\tau)&=\frac{1}{\sqrt{\pi}\Gamma}\,\mathrm{e}^{-\Delta^2}\,, &\Delta\equiv\sigma\tau\,,\quad \Gamma\equiv\frac{\gamma}{\sigma}.
 \end{align}
 Since $ \delta P _\mathrm{HOM} (0) $ does not vanish   (which is true for any photon spectral function), the HOM dip is no longer ideal.
A more  general case, not restricted to $\Gamma\gg1$, is illustrated in  Fig.~\ref{Fig4} while  general expressions for both factors in Eq.~(\ref{P12})  will be given in \emph{Methods} below.

In the geometries of Fig.~\ref{pc} two incident photons tuned into the resonance $\omega=0$ are  coming through the same channel ($\beta=\beta' =1$). As the two geometries are dual, we give the results on that of Fig.~\ref{pc}b.  In this case the nonlinearity suppresses the resonant transmission as two photons cannot simultaneously excite the TLS. Such a ``weak photon blockade" by TLS results in the nonzero probability,  $P_\mathrm{res}$, of one photon transmitted and the other reflected.  This probability is peaked for simultaneous photon arrival, $\Delta=0$. For photons with narrow Gaussian spectral function we find
\begin{align}\label{blk}
     \delta P_\mathrm{res}=P_\mathrm{res}-P^0_\mathrm{res}=\zeta_{\mathrm{bl}}=\frac{1}{\sqrt{\pi}\Gamma}
\frac{1}{1+\mathrm{e}^{ \Delta^2}}\,,
\end{align}
  In this limit the blockade is weak even in the absence of the time delay $\tau$  since the TLS re-emits photon over the time interval $\gamma^{-1} $ which is much shorter than the time interval  $\sigma^{-1}$ over which each photons arrives. However, with $\Gamma$ decreasing both the nonlinearity becomes weaker and a trivial single-photon off-resonant scattering plays a bigger role.  Thus the full anti-bunching probability has a maximum at $\Gamma\sim1$ as shown on the inset in Fig.~\ref f.

\begin{figure}[t]
\includegraphics[width=0.94\columnwidth]{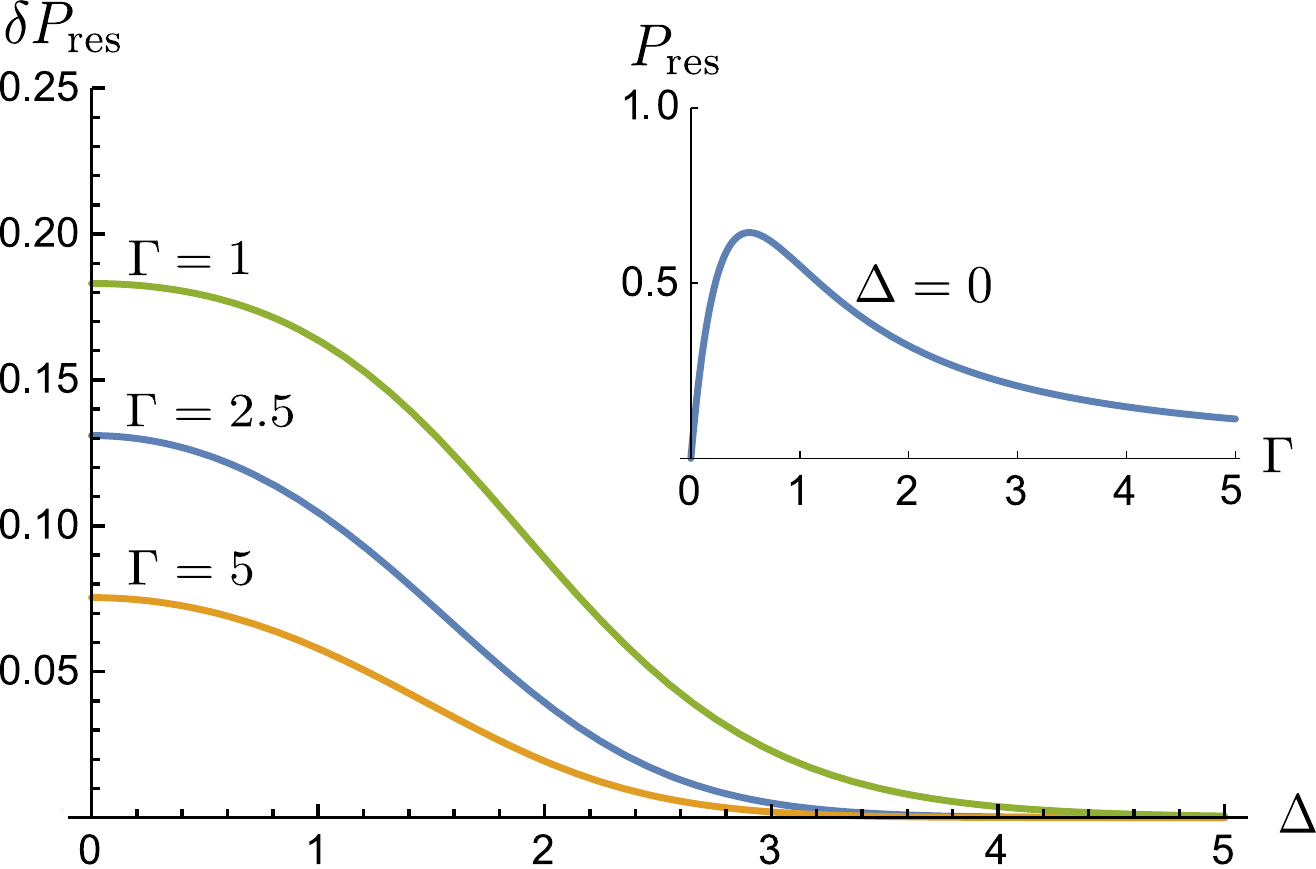}
\caption{Weak photon blockade. \label{f} Here $\delta P_\mathrm{res}$ is the nonlinearity-induced anti-bunching probability in the resonance geometry of Fig.~\ref{pc}(b), i.e.\ the probability of reflecting    one out of two photons, as a function of their relative delay $\Delta=\sigma\tau$. The inset shows a non-monotonic behaviour of the total anti-bunching probability due to a trivial off-resonance reflection at smaller $\Gamma$.
}
\end{figure}

\section{Methods}
The multi-photon scattering states for the model (\ref{H}) can be described in terms of the Bethe ansatz eigenstates, see e.g.\  Refs.~\cite{sf,yr}, each state parameterised by a string of complex quasi-momenta. The physical implication of these complex-valued quasi-momenta is the emergence of many-photon   states  weakly bound to  the atom (TLS). The asymptotic scattering states constructed out of these eigenstates will be affected by such binding.

For the two-photon scattering states we will use a more straightforward and physically transparent approach based on the scattering matrix. Note that in two-photon (or multi-photon) scattering described by the model (\ref{H}) two opposing tendencies compete: the statistical photon bunching clearly demonstrated by the dip in the coincidence counter in the idealised HOM geometry, and the nonlinearity caused by the fact that at any moment only a single photon can go through the TLS.  It is this competition that spoils the ideal antibunching in the HOM geometry   and the ideal resonant transmission (or reflection) in the geometries of Fig.~\ref{pc}.

Scattering properties of the atom-light interaction described by Hamiltonian (\ref{H}) are encoded in the $S$-matrix, Eqs.~(\ref{SM})-(\ref{sa}), which links asymptotic incoming states to the  corresponding outgoing states. Inevitable spontaneous emission means that in both the incoming and outgoing asymptotic state the TLS is  in its ground state so that in specifying the asymptotic states one needs to  refer explicitly only to their photonic part. To allow for time-resolved photons, in what follows we consider photons with a  spectral function   $g({\omega-\omega_0})$, centred around $\omega_0$ with a spectral width $\sigma$. For analytic calculations it can be chosen Gaussian,
 \begin{align*}
    g({\omega-\omega_0}) &= \sqrt{\frac{1}{\sigma\sqrt{\pi}}}\exp\left[-\frac{(\omega-\omega_0)^{2}}{2\sigma^{2}}\right],
\end{align*}
but in general all we need to know is its centre, $\omega_0$, and width, $\sigma$. As we are  considering on-shell scattering,  $\omega=q$ (in the units where the group velocity of light is $1$), we may also refer to this function as $g({q-\omega_0})$.

Then the incoming one-photon state is given by
\begin{align}\label{inq}
    \ket\beta_\mathrm{in}&= \varint \frac{\rm{d}q}{2\pi}\; g({q-\omega_0})\hat{b}^\dagger _\beta( q)\Ket{0} \,.
\end{align}
When the spectral function   centres at $ |\omega_0|=\gamma$, it corresponds to the scattering emulating the balanced beam splitter in the HOM geometry, while the zero detuning,  $\omega_0=0$, corresponds to the exact resonance (or antiresonance) in the geometry of Fig.~\ref{pc}. In the time-resolved representation the photon amplitude, which is a Fourier transform of $g({q-\omega_0})$, is a running wave, $\tilde g ({x-t})$ centred at $x=t$ of width of order $1/\sigma$.

A two-photon incoming state can be represented as
\begin{align}\label{B}
\Ket{{\beta,\beta'}}_\mathrm{in}=\int\frac{{\rm d}^2q}{(2\pi)^2}\,B_{\beta\beta'}(q,q')\,\ket{\beta,q; \beta' ,q'}
\end{align}
where $B_{\beta\beta'}(q,q')$ is a two-photon amplitude of the state $\ket{\beta,q; \beta' ,q'}\equiv\hat{b}^{\dagger}_{\beta}(q)\, \hat{b}^{\dagger}_{\beta'}(q')\,\Ket{0}$. For uncorrelated incoming photons    $B_{\beta\beta'}(q,q')$ is a direct product of their spectral functions.   For two identical photons in the HOM geometry, scattering from the TLS with time delay $\tau$, the amplitude is given by
\begin{align}\label{BB}
B_{12}(q,q')=g(q)\,g(q')\,\mathrm{e}^{iq\tau}\,,
\end{align}
while for the `weak blockade' geometry the appropriate function $B_{11}(q,q') $ differs from the above by symmetrising with respect to $q$ and $q'$.

The $q$-dependent one-photon scattering matrix is  given by
\begin{align}
\mathcal{S}_{\alpha\beta}^{kq}\equiv\Bra{\alpha, k} \hat{\mathrm{S}}\Ket{\beta, q}&=2\pi\,\delta(k-q)\,s_{\alpha\beta}(q)\,,
\end{align}
where the on-shell $q$ dependence is equivalent to the $\omega$ dependence given in Eq.~{(\ref{sa})}.
To separate statistical and interaction effects,  one relates the two-photon S-matrix with the appropriate T-matrix,  $$\hat{\mathcal{S}}=\hat{1}-i\hat{\mathcal{T}} =\hat{1}-i\hat{\mathcal{T}}^0-i\hat{\mathcal{T}}_\mathrm{int}\equiv \hat{\mathcal{S}}^0-i\hat{\mathcal{T}}_\mathrm{int},$$
where $\hat{\mathcal{S}}^0$ (or $\hat{\mathcal{T}}^0$) describes scattering of  non-interacting photons while $i\hat{\mathcal{T}}_\mathrm{int}$ represents the nonlinearity contribution. The matrix elements of $\hat{\mathcal{S}}^0$ are given by
\begin{align}\label{S0}
\!\Bra{\alpha,k; \alpha', k'}\hat{\mathcal{S}}^0\Ket{\beta, q; \beta', q'}=S_{\alpha\beta}^{kq} S_{\alpha'\beta'}^{k'q'}  + S_{\alpha\beta'}^{kq'} S_{\alpha'\beta}^{k'q} ,
\end{align}
while those of $\hat{\mathcal{T}}$  are found by accounting for the phase space restriction on the TLS occupation  by photons
\begin{align}\label{tau}
\!\!\Bra{\alpha,k; \alpha',k'}{\hat {\mathcal{T}}}\Ket{\beta,q;\beta',q'}
=-\frac{4\Gamma^2\,\Omega_{{k}}}{A_{{k}}A_{{q}}}
\,2\pi\,\delta(\varepsilon_{{k}}\!-\!\varepsilon_{{q}}),
\end{align}
where
\begin{align}
\varepsilon_{{k}}&=k+k'\,,\quad 2\xi_{{k}}=k-k'\,,\quad \notag \Omega_{{k}}=\frac{\varepsilon_{{k}}}{2}+i\gamma\,,\\
A_{{k}}&=(k+i\gamma)\,(k'+i\gamma)=\Omega^2_{{k}}-\xi^2_{{k}}\,.\label{ek}
\end{align}
 On substituting the incoming state of Eq.~({\ref{B}}), one reduces the matrix element Eq.~(\ref{tau}) to
\begin{align}
\Bra{\alpha,k; \alpha',k'}{\hat {\mathcal{T}}}\Ket{\beta,\beta'}\
=-\frac{4\Gamma^2\,\Omega_{{k}}}{A_{{k}}}\int\frac{{\rm d}\xi_{{q}}}{2\pi}\,\frac{B_{\beta\beta'} }{\Omega^2_{{k}}-\xi^2_{{q}}}\,.
\end{align}
 The probability of finding outgoing photons in the channels $\alpha$ and $\alpha'$ for the incoming state of Eq.~(\ref{B})  is
\begin{align}
P_{\alpha\alpha' |\beta\beta'}&=\frac{1}{N_{\beta\beta'}} \int\!\!\frac{{\rm d} k \, \mathrm{d}k'}{(2\pi)^2}\,\left|\Bra{\alpha,k;\alpha',k'}\,\hat{S}\, \Ket{\beta,\beta'}\right|^2\!,\notag\\
N_{\beta\beta'}&= (1+\delta_{\beta\beta'}) \int\frac{{\rm d} q\,\mathrm{d}q'}{(2\pi)^2}\,|B_{\beta\beta'}(q,q')|^2\,.\label{Paa}
\end{align}
Substituting   the matrix elements of $\hat{S}$, corresponding to the independent scattering, Eq.~(\ref{S0}), and the nonlinearity induced scattering, Eq.~(\ref{tau}),  we can decompose  the probability in each channel as the sum $P^0+\delta P$,
where $\delta P$ is the nonlinearity induced change in the probability of photon bunching ($\alpha=\alpha'$) or anti-bunching ($\alpha\ne\alpha'$). Both in the HOM and resonance geometries we are interested in the anti-bunching probabilities, $P_\mathrm{HOM}=P_{12|12}+P_{21|12}  $ and $P_\mathrm{res}=P_{12|\beta\beta}+P_{21|\beta\beta}  $.

We focus on the limit $\Gamma\equiv\gamma/\sigma\gg1$ that corresponds to almost monochromatic photons strongly scattered by the TLS at the resonance \cite{note1}. In this case the two-photon spectral function of Eq.~(\ref{BB}) is sharply peaked at $q=q'=\omega_0$ while the matrix elements of $S^0$, Eq.~(\ref{S0}), are smooth functions of wave numbers so that their dispersion can be neglected. All further analytic results are found in this approximation.

In the HOM geometry, $\beta\neq\beta'$, the noninteracting part of the probability matrix is governed by the transmission and reflection probabilities, $T=|t_{\omega_0}|^2$ and $R=|{r_{\omega_0}}| ^2$ with the scattering amplitudes given by Eq.~(\ref{sa}), and by the following coherence factor,
\begin{align}\label{nu}
\nu=N^{-1}_{12}\Re \int\!\frac{{\rm d}\varepsilon_k }{2\pi}\,\frac{{\rm d}\xi_k}{2\pi}\,\bar{B}_{12}(\varepsilon_k ,\xi_k)B_{12}(\varepsilon_k ,-\xi_k)\,,
\end{align}
 {which is easily found in the Gaussian case, using Eq.~(\ref{BB}) and (\ref{ek})  as} $\nu=\mathrm{e}^{-\Delta^2} $.
In these terms we find $P^0_\mathrm{HOM}=({T^2+R^2})-2\nu\,TR $.
        For two identical photons in the absence of the delay between them ({$\Delta=0$}) the coherence factor $\nu=1$  so that $P^0_\mathrm{HOM}=({T-R})^2$, which vanishes when the central photon frequency $\omega_0=\gamma$ (the point where the model emulates the balanced beam splitter).

        The nonlinearity-induced correction to $P_\mathrm{HOM}$ is found, using Eq.~(\ref{tau}) and (\ref{Paa}),   as
                                 \begin{align}\label{HOMR}
                                     \delta {P}_{\mathrm{HOM}}  = -2\zeta_{12} \,T^2\,(1-4R)
                                 \end{align}
where
\begin{align}\label{zeta}
\zeta_{\beta\beta'}=N_{\beta\beta'}^{-1}\,\int\frac{{\rm d}\varepsilon_k}{2\pi\gamma}\,\left|\int\frac{{\rm d}\xi_k}{2\pi}\,B_{\beta\beta'}(\varepsilon_k ,\xi_k)\right|^2\,.
\end{align}
Calculating this integral for the Gaussian spectral function results in $\zeta_{12}$ given by Eq.~(\ref{HOMres}).

In the resonance geometry, where both photons are incoming from the same channel ($\beta=\beta'=1$), one finds $P^0_\mathrm{res}=2TR$
provided that $\Gamma\gg1$. In this limit in the geometry of Fig.~\ref{pc}(b)
  independent photons are   both  transmitted at the exact resonance, $T=1$ and $R=1-T=0$.
The   nonlinearity correction is found similar to that for the HOM geometry as
                                 \begin{align}\label{resR}
                                     \delta {P}_{\mathrm{res}}  =   \zeta_{11} \,T^2\,(1-4R)
                                 \end{align}
where $\zeta_{ {11 }}$ is found from Eq.~(\ref{zeta}) and (\ref{Paa}).   In the exact resonance $P^0_\mathrm{res}=0$ and the anti-bunching probability, describing the weak photon blockade,   reduces to $\delta P_\mathrm{res}=\zeta_{11} \equiv\zeta_\mathrm{bl} $. The expression for $\zeta_\mathrm{bl}$  for the Gaussian case   is  given by Eq.~(\ref{blk}) in Results.

In the opposite limit, $\Gamma\ll1$, the resonance effects are weak since the spectral width $\sigma$ greatly exceeds the resonance width $\gamma$. This trivial case is of little interest but in Fig.~\ref{f} we give numerical results for the weak photon blockade for intermediate values of larger $\Gamma$ up to $\Gamma=1$.  On the inset in this figure we show the full reflection probability  of one out of two simultaneously arriving photons. Its maximum at $\Gamma\sim1$ is higher than $\zeta_\mathrm{bl}$ due to a trivial off-resonance reflection. Note that this result (i.e.\ in the absence of the delay) is in agreement with those earlier obtained by Zheng et al\cite{bar}.

\section{Summary}
In this work we have analysed the role of nonlinearity in two-photon propagation in Hong-Ou-Mandel geometry and in the resonance geometry in photonic crystals. In the ideal HOM geometry two identical photons simultaneously arriving at a balanced beam splitter through different incoming ports are bunched together to leave the system through the same outgoing port. The HOM interferometer is thus used for measuring the degree of entanglement in biphotons produced by SPDC. Here we argue that the presence of the nonlinearity in the beam splitter results in partial anti-bunching of the photons thus potentially obscuring such a measurement. We model the nonlinearity by a two-level system (which at the same time emulates a balanced beam splitter) and calculate the  anti-bunching probability induced by this nonlinearity. Our results are given by Eq.~(\ref{HOMR}) and (\ref{zeta}) and illustrated in Fig.~\ref{Fig4}.

We use the same model to illustrate the ``weak photon blockade" of resonant two-photon propagation through a TLS.  The blockade is due to the impossibility of double-occupancy of the TLS, and it is always weak due to inevitable spontaneous emission. However, exactly at the resonance this leads to a finite anti-bunching probability, as illustrated in Fig.~\ref f  while two independent photons would be either both resonantly reflected (Fig.~\ref{pc}a) or both resonantly transmitted (Fig.~\ref{pc}b).  We expect this effect to be considerably more pronounced for multiple photon propagation which requires further analysis.

\acknowledgments
{We gratefully acknowledge support from the Leverhulme Trust via Grant No. RPG-380}


\begin{thebibliography}{10}

\bibitem{HOM1987}
C.~K. Hong, Z.~Y. Ou, and L. Mandel, Measurement of subpicosecond time
  intervals between 2 photons by interference, {\em Phys. Rev. Lett.} {\bf 59},
   2044  (1987).

\bibitem{Mandel1995}
L. Mandel and E. Wolf, {\em Optical Coherence and Quantum Optics} (CUP,
  Cambridge, UK, 1995).

\bibitem{Ou1988}
Z.~Y. Ou and L. Mandel, Violation of {B}ells-inequality and classical
  probability in a 2-photon correlation experiment, {\em Phys. Rev. Lett.} {\bf
  61},  50  (1988).

\bibitem{Pan2001}
J. Pan {\it et~al.}, Experimental demonstration of four-photon entanglement and
  high-fidelity teleportation, {\em Phys. Rev. Lett.} {\bf 86},  4435  (2001).

\bibitem{Kuzucu:05}
O. Kuzucu {\it et~al.}, Two-photon coincident-frequency entanglement via
  extended phase matching, {\em Phys. Rev. Lett.} {\bf 94},  083601  (2005).

\bibitem{Mosley2008}
P.~J. Mosley {\it et~al.}, Heralded generation of ultrafast single photons in
  pure quantum states, {\em Phys. Rev. Lett.} {\bf 100},  133601  (2008).

\bibitem{Eckstein:08}
A. Eckstein and C. Silberhorn, Broadband frequency mode entanglement in
  waveguided parametric downconversion, {\em Opt. Lett.} {\bf 33},  1825
  (2008).

\bibitem{Nasr2008}
M.~B. Nasr {\it et~al.}, Ultrabroadband biphotons generated via chirped
  quasi-phase-matched optical parametric down-conversion, {\em Phys. Rev.
  Lett.} {\bf 100},  183601  (2008).

\bibitem{Ray:11}
I.~ M. Mirza and S.J. van Enk, {\em Optics Communications} {\bf 343}, 172 (2015);
\,M.~R. Ray and S.~J. van Enk, Verifying entanglement in the Hong-Ou-Mandel dip,
  {\em Phys. Rev. A} {\bf 83},  042318  (2011).

\bibitem{Douce:13}
T. Douce {\it et~al.}, Direct measurement of the biphoton Wigner function
  through two-photon interference, {\em Sci. Rep.} {\bf 3},  3530  (2013).

\bibitem{Jin2014}
H. Jin {\it et~al.}, On-Chip Generation and Manipulation of Entangled Photons
  Based on Reconfigurable Lithium-Niobate Waveguide Circuits, {\em Phys. Rev.
  Lett.} {\bf 113},  103601  (2014).

\bibitem{LYY:08}
I.~V. Lerner, V.~I. Yudson, and I.~V. Yurkevich, Quantum wire hybridized with a
  single-level impurity, {\em Phys. Rev. Lett.} {\bf 100},  256805  (2008).

\bibitem{note1}
let us stress that for a typical TLS (with $\gamma^{-1}\sim 10^{-8}\,s $) such
  `almost monochromatic' photons can still be sharply time-resolved for all
  practical purposes as $\sigma^{-1} $ can be much shorter than, e.g., the time
  resolution of a coincidence counter. (unpublished).

\bibitem{sf}
J.-T. Shen and S. Fan, Strongly correlated two-photon transport in a
  one-dimensional waveguide coupled to a two-level system, {\em Phys. Rev.
  Lett.} {\bf 98},  153003  (2007).

\bibitem{yr}
V. Yudson and P. Reineker, Multiphoton scattering in a one-dimensional
  waveguide with resonant atoms, {\em Phys. Rev. A} {\bf 78},  052713  (2008).

\bibitem{bar}
H. Zheng, D. Gauthier, and H. Baranger, Waveguide QED: Many-body bound-state
  effects in coherent and Fock-state scattering from a two-level system, {\em
  Phys. Rev. A} {\bf 82},  063816  (2010).

\end{thebibliography}

\end{document}